\documentclass[a4paper]{jpconf}
\usepackage{graphicx,amsmath,dcolumn,bm}
\usepackage{wrapfig,epsfig}
\usepackage{epic,eepic}

\begin{document}

\title{Clustering structure of nuclei in deep inelastic processes}

\author{M. Hirai$^{\, a}$, S. Kumano$^{\, b,c,d}$, K. Saito$^{\, d,e}$, 
        T. Watanabe$^{\, e}$}
\address{$^a$ Nippon Institute of Technology, Saitama 345-8501, Japan \\
         $^b$ KEK Theory Center,
             Institute of Particle and Nuclear Studies,
             KEK \\
             \ \ and Department of Particle and Nuclear Studies,
                 Graduate University for Advanced Studies \\
             \ \ 1-1, Ooho, Tsukuba, Ibaraki, 305-0801, Japan \\
          $^c$ Theory Group, Particle and Nuclear Physics Division, 
               J-PARC Center \\
             \ \ 203-1, Shirakata, Tokai, Ibaraki, 319-1106, Japan \\
          $^d$ J-PARC Branch, KEK Theory Center,
             Institute of Particle and Nuclear Studies, KEK \\
             \ \ 203-1, Shirakata, Tokai, Ibaraki, 319-1106, Japan \\
          $^e$ Department of Physics, Faculty of Science and Technology,
             Tokyo University of Science \\ 
             \ \ 2641, Yamazaki, Noda, Chiba, 278-8510, Japan}

\ead{shunzo.kumano@kek.jp}

\begin{abstract}
A clustering aspect is explained for the $^9$Be nucleus in 
charged-lepton deep inelastic scattering. Nuclear modifications 
of the structure function $F_2$ are studied by the ratio
$R_{\rm EMC} = F_2^A /F_2^D$,
where $A$ and $D$ are a nucleus and the deuteron, respectively.
In a JLab experiment, an unexpectedly large nuclear modification 
slope $|dR_{\rm EMC}/dx|$ was found for $^9$Be, which could be related 
to its clustering structure. We investigated a mean conventional part 
of a nuclear structure function $F_2^A$ by a convolution description 
with nucleon momentum distributions calculated by antisymmetrized
(or fermionic) molecular dynamics (AMD) and also by a simple shell model. 
We found that clustering effects are small in the conventional part, 
so that the JLab result could be associated with 
an internal nucleon modification or a short-range nuclear correlation
which is caused by high densities due to cluster formation.
\end{abstract}

\vspace{-0.3cm}

\section{Introduction}

Nuclear modifications of the structure function $F_2$ are known
as the EMC effect \cite{sumemc}, and such effects are shown by the ratio 
$R_{\rm EMC} (x) = F_2^A (x)/F_2^D (x)$, where $A$ and $D$ 
indicate a nucleus and the deuteron, respectively, and
$x$ is the Bjorken scaling variable. 
 Various mechanisms contribute to the modifications.
At large $x >0.7$, nucleon Fermi motion in a nucleus gives rise
to increase of the ratio $R_{\rm EMC} (x)$.
At medium $x$ ($0.3<x<0.7$), nuclear binding is the major source
of nuclear effects. In addition, modifications of internal
nucleon structure could contribute. At small $x <0.05$,
the nuclear shadowing effects become prominent.
The nuclear structure functions have been measured from small $x$ to 
large $x$ for many nuclei, and the nuclear modifications have been 
determined by global analyses of experimental data \cite{npdfs}.

Nuclei are basically described by a shell model, 
where nucleons are assumed to move in an average central potential
created by interactions of all the nucleons in a nucleus.
However, cluster structure in a nucleus is not easily described
by the shell model with a limited number of shells.
We know that some nuclei in the region of the mass number $A \sim 10$
have clustering configurations. It is interesting to study whether
such clustering features appear in the nuclear structure functions $F_2^A$.

\begin{wrapfigure}[12]{r}{0.36\textwidth}
  \vspace{-0.1cm}
  \begin{center}
      \epsfig{file=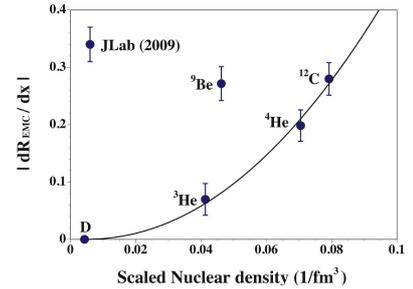,width=0.32\textwidth} 
  \end{center}
\vspace{-0.6cm}
\caption{Nuclear modification slope $| d(F_2^A/F_2^D)/dx |$.}
\label{fig:slope-jlab}
\end{wrapfigure}

In 2009, an anomalously large nuclear effect was reported 
by the JLab measurement on the beryllium-9 structure function $F_2$ 
as shown in Fig. 1 \cite{jlab-be9}, where
the slope is shown as a functions of the scaled average nuclear density.
The scaled means that the factor $(A-1)/A$ is multiplied for
removing the struck nucleon. Although other data
are on the smooth curve, the beryllium-9 data is 
much larger than the one expected from its average nuclear density. 

It is known that the beryllium-9 has a clustering configuration 
with two $\alpha$ nuclei and surrounding neutron clouds.
Therefore, such an unexpected nuclear modification
could come from the cluster formation in the nucleus. 
We investigated such a possibility by using a convolution
description for the structure function $F_2^A$
with nucleon momentum distributions
calculated by antisymmetrized molecular dynamics (AMD) and also 
by a simple shell model for comparison \cite{cluster-dis}.
Then, possible reasons are suggested for the large slope
of $^9$Be due to the cluster formation in the nucleus.

\section{Clustering effects in nuclear structure functions}

\begin{wrapfigure}[8]{r}{0.36\textwidth}
  \vspace{-1.0cm}
  \begin{center}
      \epsfig{file=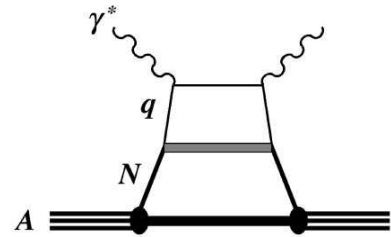,width=0.32\textwidth} 
  \end{center}
\vspace{-0.5cm}
\caption{Convolution for $F_2^A$.}
\label{fig:convolution}
\end{wrapfigure}

The standard method for describing the nuclear structure functions 
$F_2^A$ is to use the convolution description \cite{ek03} 
for describing a mean conventional part.
The structure function $F_2^A$ is given by the nucleonic one $F_2^N$
convoluted with a nucleon momentum distribution in the nucleus $f(y)$
\cite{sumemc,ek03} as shown in Fig. \ref{fig:convolution}:
\begin{align}
F_{2}^A (x, Q^2) & = \int_x^A dy \, f(y) \, F_{2}^N (x/y, Q^2) , 
\nonumber \\
f(y) & =  \frac{1}{A} \int d^3 p_N
     \, y \, \delta \left( y - \frac{p_N \cdot q}{M_N \nu} \right) 
     | \phi (\vec p_N) |^2 ,
\label{eqn:w-convolution}
\end{align}
where $y$ is the momentum fraction 
$ y   =  M_A \, p_N \cdot q /(M_N \, p_A \cdot q)
  \simeq A \, p_N^+ /(p_A^+) $
with a light-cone momentum $p^+$.

The momentum density $| \phi (\vec p_N) |^2$  is calculated
by two methods, 
antisymmetrized molecular dynamics (AMD) \cite{amd-intro} 
or fermionic molecular dynamics (FMD) \cite{fmd-intro},
and a simple shell model for comparison.
The AMD and FMD are essentially the same, so that we use
the notation AMD hereafter. The AMD is a variational method, 
and its advantage is that there is 
no a priori assumption on nuclear structure whether
it is a shell or cluster-like configuration.
A nuclear wave function is given by the Slater determinant
of single-particle wave functions
\begin{align}
\left | \Phi (\vec r_1, \vec r_2, \cdot\cdot\cdot, \vec r_A ) \right >
   & = \frac{1}{\sqrt{A!}}
            \text{det} [ \varphi_1 (\vec r_1), \varphi_2 (\vec r_2), 
                                 \cdot\cdot\cdot, \varphi_A (\vec r_A) ] .
\end{align}
Here, a nucleon is described by the single-particle wave function
$
\varphi_i (\vec r_j) = \phi_i (\vec r_j) \, \chi_i \, \tau_i ,
$
where $\chi_i$ and $\tau_i$ indicate spin and isospin states,
respectively. The function $\phi_i (\vec r_j)$ is 
the space part of the wave function, and it is assumed to be
given by the Gaussian functional form:
\begin{equation}
\phi_i (\vec r_j) = \left ( \frac{2 \nu}{\pi} \right )^{3/4}
        \exp \left [ - \nu 
           \left ( \vec r_j -\frac{\vec Z_i}{\sqrt{\nu}} \right ) ^2
             \right ] ,
\end{equation}
where $\nu$ is a parameter to express the extent of the wave packet.
The center of the wave packet is given by $\vec Z_i/\sqrt{\nu}$,
where $\vec Z_i$  is a complex variational parameter.
The variational parameters are then determined
by minimizing the system energy with a frictional-cooling method.
We use effective $NN$ interactions which could describe gross
properties of nuclei:
\begin{align}
\text{2-body:} \ \ 
     & V_2 = (1-m -m P_\sigma P_\tau)  
          \left [  v_{21} e^{- (r/r_{21})^2} 
                 + v_{22} e^{- (r/r_{22})^2}  \right ] ,
\nonumber \\
\text{3-body:} \ \ 
     & V_3 =  v_3 \, \delta^3 (\vec r_1 - \vec r_2 ) 
                  \, \delta^3 (\vec r_2 - \vec r_3 ) ,
\nonumber \\
\text{LS:} \ \ 
     & V_{LS} = v_{LS} \left [ e^{- (r/r_{LS1})^2} 
                  - e^{- (r/r_{LS2})^2} \right ] 
        P(^3 O) \,  \vec L \cdot \vec S ,
\label{eqn:amd-interactions}
\end{align}
where $m$, $v_{21}$, $v_{22}$, $r_{21}$, $r_{22}$, 
$v_3$, $v_{LS}$, $r_{LS1}$, and $r_{LS2}$ are constants. 

The momentum distributions of Eq. (1) are also calculated 
in a simple shell model, 
where the potential is the harmonic-oscillator type 
($M_N \omega^2 r^2/2$),
for clarifying cluster effects by comparing two results.
Its wave function is separated into radial- and angular-dependent parts:
$
\psi_{n \ell m} (r, \theta, \phi) = R_{n \ell} (r) Y_{\ell m} (\theta, \phi)
$
where $r$, $\theta$, and $\phi$ are spherical coordinates,
and $n$, $\ell$, and $m$ are radial, azimuthal, and magnetic quantum numbers,
respectively. 
The function $Y_{\ell m} (\theta, \phi)$ is the spherical harmonics, and
the radial wave function is given by
\begin{equation}
\! 
R_{n \ell} (r) = \sqrt{\frac{2 \kappa^{2\ell+3}(n-1)!}
                            {[\Gamma(n+\ell+1/2)]^3}}
                  r^\ell e^{-\frac{1}{2}\kappa^2 r^2} 
                  L_{n-1}^{\ell+1/2} (\kappa^2 r^2) ,
\end{equation}
where $L_{n-1}^{\ell+1/2} (x)$ is the Laguerre polynomial, and $\kappa$ 
is defined by $\kappa \equiv \sqrt{M_N \omega}$.

\begin{wrapfigure}[12]{r}{0.40\textwidth}
  \vspace{-0.4cm}
  \begin{center}
      \epsfig{file=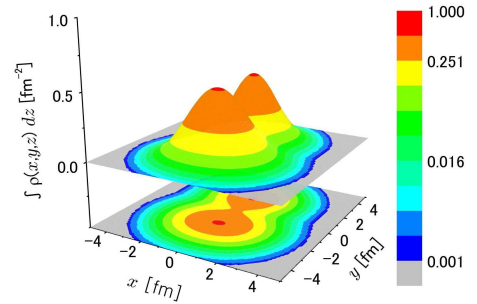,width=0.38\textwidth} 
  \end{center}
\vspace{-0.5cm}
\caption{Density of $^9$Be by AMD.}
\label{fig:amd-9be}
\end{wrapfigure}

As a result, we obtain the two-dimensional spacial density 
of the $^9$Be nucleus in Fig. 3.
It obviously indicates the existence of two $\alpha$ clusters,
together with surrounding neutron clouds.
In a simple shell model, such clusters do not appear,
and the density is a monotonic distribution.
If the averages are taken over polar and azimuthal angles
$\theta$ and $\phi$, the distributions becomes the ones in Fig. 4. 
After the averages, the cluster structure is no longer apparent.
Because the nucleons are confined mainly in two separate regions,
the radial distribution is shifted toward
large-$r$ region for $^9$Be in Fig. 4 as shown by the solid
curve in comparison with the dashed one.

\begin{figure}[b]
\begin{center}
   \includegraphics[width=0.32\textwidth]{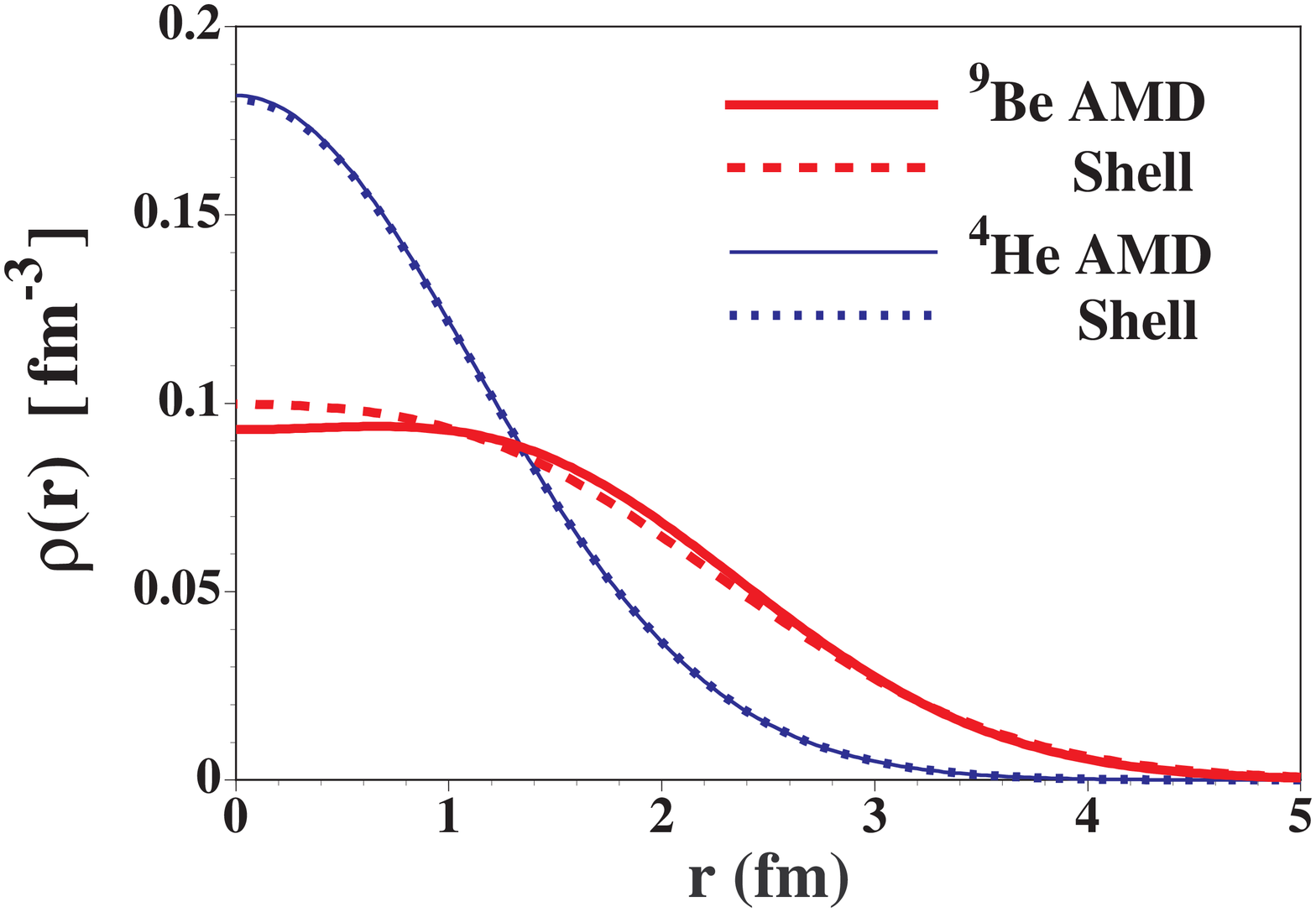}
   \hspace{2.2cm}
   \includegraphics[width=0.32\textwidth]{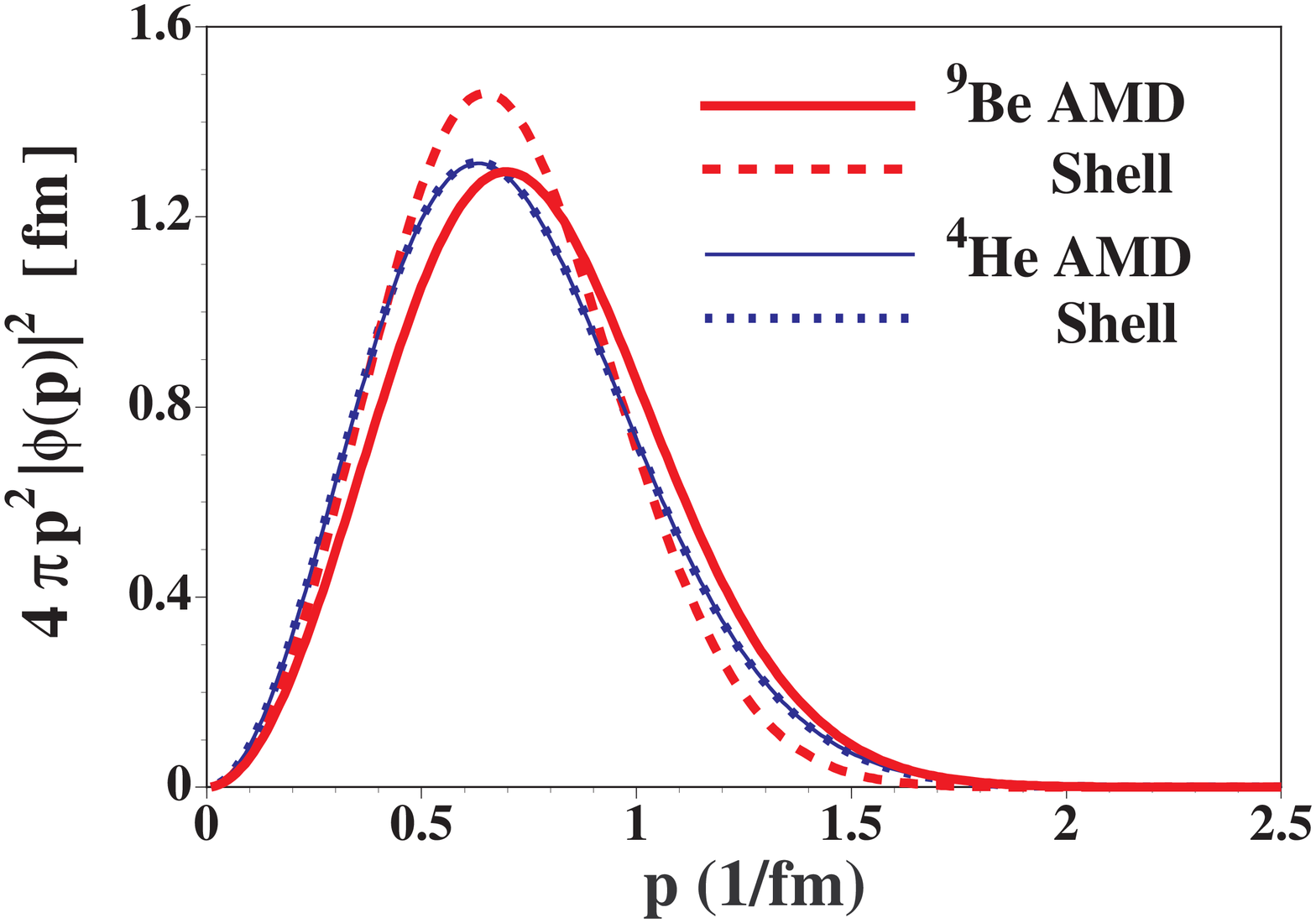}
\end{center}
\vspace{-0.3cm}
\noindent
{
 \hspace{1.3cm}
 {\bf Figure 4.} Spatial distributions.
 \hspace{1.4cm}
 {\bf Figure 5.} Momentum distributions.
 }
\vspace{-0.2cm}
\end{figure}

The momentum distributions in Fig. 5 are calculated from 
the spacial distributions.
Both AMD and shell-model results are shown for $^4$He and $^9$Be.
The momentum distribution of the AMD is shifted toward 
the high-momentum region in $^9$Be, which
is caused by the fact that the dense clusters are formed 
within the $^9$Be nucleus. If nucleons are confined
in the small space regions of the clusters as indicated
by the AMD, it leads to an increase of high momentum components.

\begin{wrapfigure}[12]{r}{0.36\textwidth}
\setcounter{figure}{5}
  \vspace{-0.05cm}
  \begin{center}
      \epsfig{file=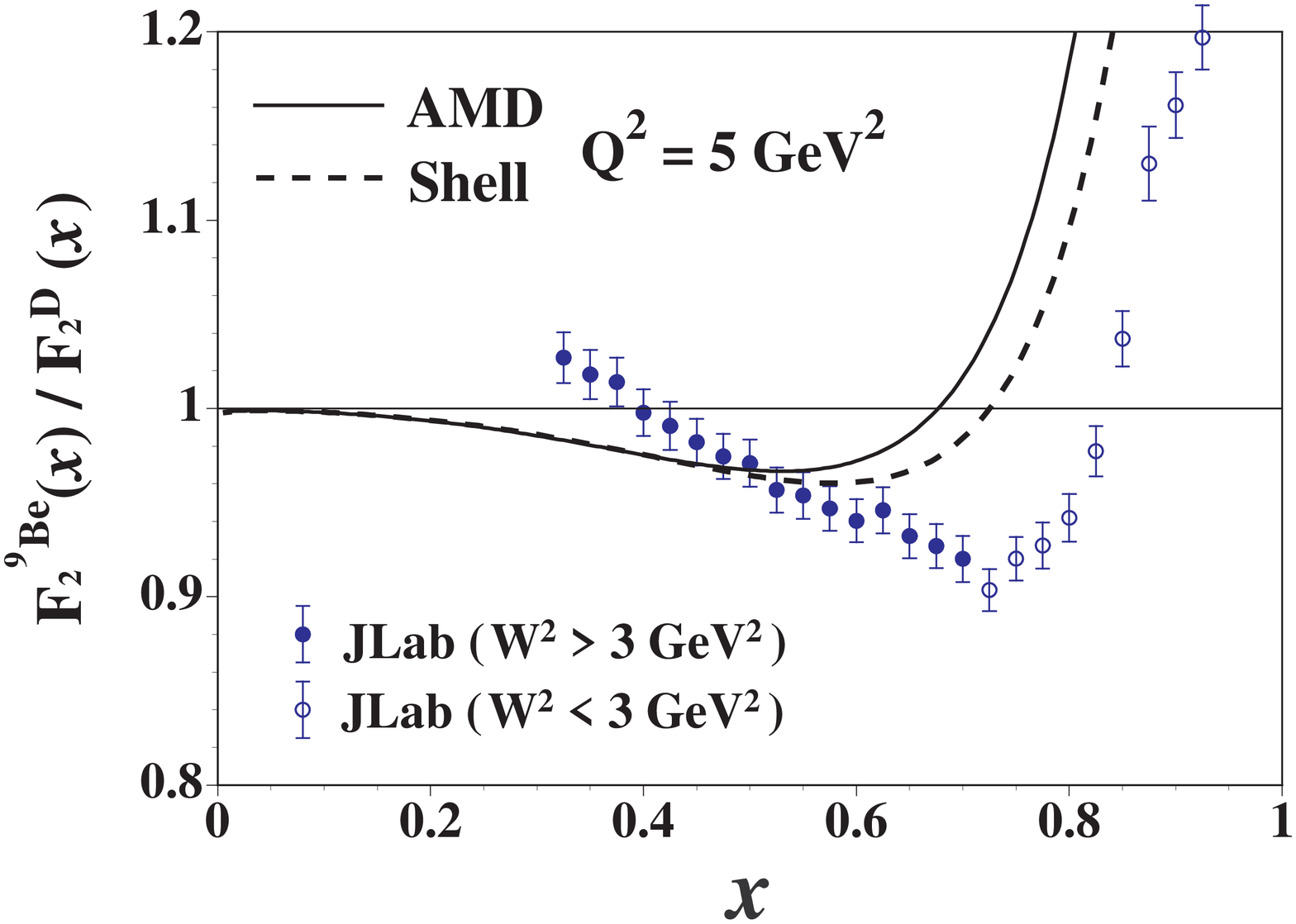,width=0.32\textwidth} 
  \end{center}
\vspace{-0.6cm}
\caption{$F_2^{\, ^9 Be} / F_2^D$ by shell 
         and AMD models.}
\label{fig:f2-ratio}
\end{wrapfigure}

The modifications of the momentum distributions should affect
the nuclear structure function in Eq.\,(1), and the results are
shown for the structure function $F_2$
in Fig. 6. There are two theoretical curves by the AMD and
shell model. In general, the theoretical ratios are consistent
with the data in the sense that the ratio decreases at medium $x$ 
and it increases at large $x$. These decrease and increase
are caused by the nuclear binding and the nucleon's Fermi motion,
respectively. However, the simple convolution model 
is not sufficient to explain the experimental data.
The differences should be attributed to the effects of
short-range nucleon-nucleon correlations and modifications
of internal nucleon structure. 
The correlation effects change the theoretical ratios toward
the experimental data at $x=0.6-0.8$. Since we are interested
in an order-of-magnitude estimate of clustering effects, we did not
step into such details in this work.
In any case, the differences of the two curves, namely 
the clustering effects, are not large and they are of
the same order of experimental errors. 
In Fig. \ref{fig:f2-ratio}, we found that the clustering
effects are rather small in the mean conventional part
of $F_2^A$. Therefore, there should be other sources 
for the large slope of $^9$Be.
We consider that the nuclear structure functions consist 
of the mean conventional part and the remaining one depending 
on the maximum local density:
\vspace{-0.05cm}
\begin{equation}
F_2^A = \text{(mean part)} 
       +\text{(part created by large densities
               due to cluster formation)}.
\end{equation}
\vspace{-0.05cm}\hspace{-0.15cm}
\noindent
The first part was calculated by the convolution model
as shown in Fig. 6, and the second part seems to 
be important for explaining the JLab $^9$Be result.
The second term is associated with the modification
of internal nucleon structure and/or a short-range
correlation caused by the cluster formation.

\begin{wrapfigure}[13]{r}{0.36\textwidth}
\setcounter{figure}{6}
  \vspace{-0.6cm}
  \begin{center}
      \epsfig{file=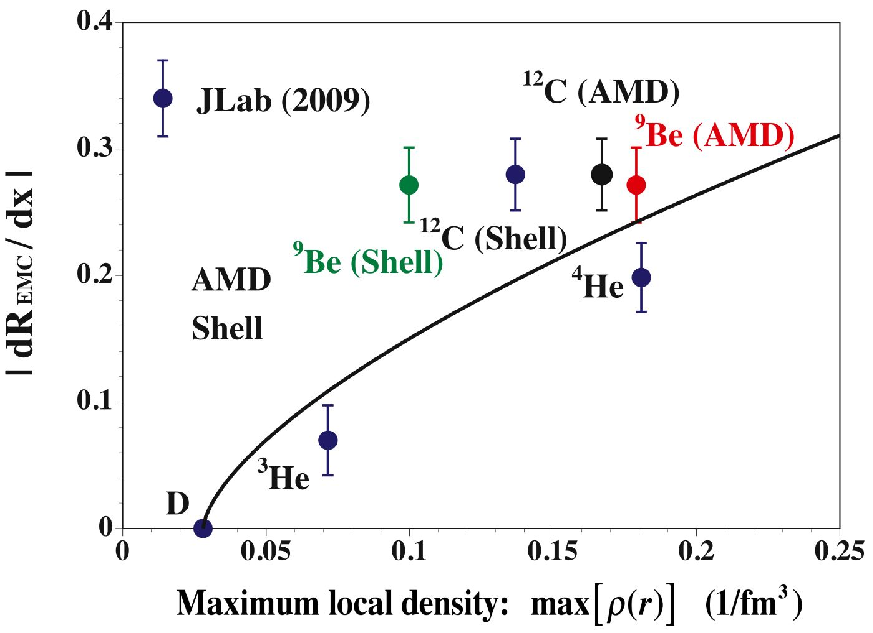,width=0.32\textwidth} 
  \end{center}
\vspace{-0.6cm}
\caption{Nuclear modification slopes 
 shown by maximum local densities.}
\label{fig:slopes-local}
\end{wrapfigure}

Although we do not step into detailed calculations in this work,
we would like to find whether such an idea works. If the physics
mechanism behind the JLab $^9$Be result is related to the cluster formation,
observables could be interpreted by the cluster densities.
For such estimates, we calculated maximum local densities 
by both AMD and shell models, rather than the average nuclear
densities. In Fig. 7, the JLab data are plotted by the maximum local 
densities calculated in the AMD and shell model.
The maximum densities are almost the same in the nuclei,
$^3$He, $^4$He, and $^{12}$C, whereas the AMD density is much
different from the shell-model one for $^9$Be.
According to the shell model with a limited number of shells,
the cluster formation, namely the large densities, cannot be
described, and it leads to a small maximum density of $^9$Be in Fig. 7.
It is also interesting to find that there is also a small
cluster effect in the $^{12}$C.
From Fig. 7, we find that all the data are on the smooth curve
if the density is calculated by the AMD for $^9$Be; however,
it is far away from the curve if it is evaluated by the shell model.
The figure indicates that the large nuclear modification slope
$|dR_{\rm EMC}/dx|$ of $^9$Be should be related to 
the high densities created by the cluster formation.

The JLab measurement could be the first result on 
a clustering aspect of a nucleus 
in high-energy nuclear reactions. It could be associated with 
internal nucleon modifications and/or short-range nuclear correlations, 
which are caused by the high densities due to the cluster configuration.
In future, there will be measurement for other nuclei, such as
$^6$Li, $^7$Li, $^{10}$B, and $^{11}$B \cite{jlab-exp},
so that much detailed information will become available.

\section{Summary}

The large nuclear modification slope of the structure function $F_2$
was found for $^9$Be, which has a typical cluster configuration.
We showed that the clustering effects are rather small
in the mean conventional part by using the convolution model
for the structure functions $F_2^A$ with the momentum distributions
calculated in the AMD and also in the shell model.
The JLab result should be associated with the high-density 
formation due to the clusters in the $^9$Be nucleus.
We showed it by plotting the JLab data as a function
of the maximum local density of a nucleus. We have not 
clarified the physics origin behind its result. However,
it is likely to be modifications of internal nucleon
structure in the nuclear medium
and/or short-range nucleon-nucleon correlations in the dense clusters. 
There is an approved experiment at JLab to measure
the EMC effects in the mass region $A \sim 10$, so that
much details will be investigated in the near future
for possible clustering effects in deep inelastic
scattering (DIS). 
The cluster physics has been investigated mainly in low energy
nuclear reactions; however, DIS processes could
shed light on new aspects in both cluster physics and 
nuclear structure functions.
Cluster physicists have an opportunity to play a major role
in high-energy hadron physics.

\section*{Acknowledgements}
This work was supported by the MEXT KAKENHI Grant Number 25105010.




\section*{References}
\begin{thebibliography}{9}
\bibitem{sumemc}  D. F. Geesaman, K. Saito, and A. W. Thomas,
                        Ann. Rev. Nucl. Part. Sci. {\bf 45}, 337 (1995).
\bibitem{npdfs} M. Hirai, S. Kumano, and M. Miyama,
                       Phys. Rev. D {\bf 64}, 034003 (2001); 
                M. Hirai, S. Kumano, and T.-H. Nagai,
                       Phys. Rev. C {\bf 70}, 044905 (2004); 
                                    {\bf 76}, 065207 (2007). 
\bibitem{jlab-be9} J. Seely {\it et al.},  
                        Phys. Rev. Lett. {\bf 103}, 202301 (2009).
\bibitem{cluster-dis} M. Hirai, S. Kumano, K. Saito, and T. Watanabe,
                        Phys. Rev. C {\bf 83}, 035202 (2011).
\bibitem{ek03}  M. Ericson and S. Kumano, 
                        Phys. Rev. C {\bf 67}, 022201 (2003). 
\bibitem{amd-intro} Y. Kanada-En'yo, M. Kimura, and H. Horiuchi,
                        C. R. Physique {\bf 4}, 497 (2003).
\bibitem{fmd-intro} H. Feldmeier and J. Schnack, 
                        Rev. Mod. Phys. {\bf 72}, 655 (2000).                         
\bibitem{jlab-exp} Jefferson Lab PAC-35 proposal, 
                        PR12-10-008, J. Arrington {\it et al.} (2009).
\end{thebibliography}
\end{document}